\documentclass[apj]{emulateapj-rtx4}

\usepackage{amssymb}
\usepackage{soul}
\usepackage{graphicx}
\usepackage{amsmath}
\usepackage[usenames]{color} 
\usepackage{ulem}
\usepackage{lscape} 
\usepackage{rotating}
\def\mod{\mathop\mathrm{mod}\nolimits}

\newcommand\plotthree[3]{{%
 \typeout{Plotthree included the files #1 #2 #3}
 \centering
 \leavevmode
 \columnwidth=.32\columnwidth
 \includegraphics[width={\columnwidth}]{#1}%
 \hfil
 \includegraphics[width={\columnwidth}]{#2}%
 \hfil
 \includegraphics[width={\columnwidth}]{#3}%
}}%

\def\gv{2002\,GV\ensuremath{_{31}}}
\def\wg{2010\,WG\ensuremath{_{9}}}
\def\jj{(278361)\,2007\,JJ\ensuremath{_{43}}}
\def\jjs{2007\,JJ\ensuremath{_{43}}}
\def\ortiz{2007\,OR\ensuremath{_{10}}}
\def\ortizlong{(225088) 2007\,OR\ensuremath{_{10}}}
\def\tiunit{{\rm J}{\rm m}^{-2}{\rm K}^{-1}{\rm s}^{-1/2}}

\def\hvmag{2.34}\def\hverr{0.05}
\def\hrmag{1.49}\def\hrerr{0.05}

\def\bflux{2.52}\def\berr{1.20}
\def\gflux{5.68}\def\gerr{1.47}
\def\rflux{6.71}\def\rerr{2.03}

\begin{document}
\sloppy

\title{Large size and slow rotation of the trans-Neptunian object
\ortizlong{} discovered from Herschel and K2 observations}
\shorttitle{Kepler/Herschel observations of \ortiz{}}

\author{Andr\'as P\'al\altaffilmark{1,2}}
\email{apal@szofi.net}
\author{Csaba Kiss\altaffilmark{1}}
\author{Thomas G. M\"uller\altaffilmark{3}}
\author{L\'aszl\'o Moln\'ar\altaffilmark{1}}
\author{R\'obert Szab\'o\altaffilmark{1}}
\author{Gyula M.~Szab\'o\altaffilmark{4,5,1}}
\author{Kriszti\'an S\'arneczky\altaffilmark{1,5}}
\author{L\'aszl\'o L.~Kiss\altaffilmark{1,5,6}}
\altaffiltext{1}{Konkoly Observatory, Research Centre for Astronomy and Earth Sciences, Hungarian Academy of Sciences, H-1121 Budapest, Konkoly Thege Mikl\'os \'ut 15-17, Hungary}
\altaffiltext{2}{E\"otv\"os Lor\'and Tudom\'anyegyetem, H-1117 P\'azm\'any P\'eter s\'et\'any 1/A, Budapest, Hungary}
\altaffiltext{3}{Max-Planck-Institut f\"ur extraterrestrische Physik, Postfach 1312, Giessenbachstr., 85741 Garching, Germany}
\altaffiltext{4}{ELTE Gothard Astrophysical Observatory, H-9704 Szombathely, Szent Imre herceg \'ut 112, Hungary}
\altaffiltext{5}{Gothard-Lend\"ulet Research Team, H-9704 Szombathely, Szent Imre herceg \'ut 112, Hungary}
\altaffiltext{6}{Sydney Institute for Astronomy, School of Physics A28, University of Sydney, NSW 2006, Australia}

\begin{abstract}

We present the first comprehensive thermal and rotational analysis of
the second most distant  trans-Neptunian object \ortizlong{}. We
combined optical 
light curves provided by the {\it Kepler} space telescope -- K2 extended
mission and thermal infrared data provided by the {\it Herschel} Space
Observatory.
We found that \ortizlong{} is likely to be larger and darker
than derived by earlier studies: we obtained a diameter of
$d=1535^{+75}_{-225}\,{\rm km}$ which places \ortizlong{} in the biggest
top three trans-Neptunian objects. The corresponding visual 
geometric albedo is $p_V=0.089^{+0.031}_{-0.009}$. The light curve
analysis revealed a slow rotation rate of
$P_{\rm rot}=44.81\pm0.37\,{\rm h}$, superseded by a very few objects only.
The most likely light-curve solution is double-peaked with a slight asymmetry,
however, we cannot safely rule out the possibility of having
a rotation period of $P_{\rm rot}=22.40\pm0.18\,{\rm h}$ which corresponds
to a single-peaked solution.
Due to the size and slow rotation, the shape of the object should be a
MacLaurin ellipsoid, so the light variation should be caused by surface
inhomogeneities. Its newly derived larger diameter also implies larger surface 
gravity and a more likely retention of volatiles --
${\rm CH}_4$, ${\rm CO}$ and ${\rm N}_2$ -- on the surface.

\end{abstract}


\section{Introduction}
\label{sec:introduction}

Trans-Neptunian objects (TNOs) are known as the most pristine types
of bodies orbiting in the Solar System. Extending our knowledge of
these objects helps us to understand both the formation of our 
planetary system and the interpretation of observational data regarding to
circumstellar material or debris disks of other stars. \ortizlong{}, discovered by 
\cite{schwamb2009}, is the second most distant known TNO to date,
following Eris: the current heliocentric distance of this object exceeds
$87\,{\rm AU}$ and still moving further away up to its aphelion in year 2130 at 
$\sim 100.7\,{\rm AU}$. Its orbital eccentricity is high ($e\approx 0.51$), 
so upon perihelion, it comes nearly as close as Neptune. In addition,
\ortiz{} is likely to be in the $3:10$ mean motion
resonance with Neptune\footnote{http://www.boulder.swri.edu/\~{ }buie/kbo/astrom/225088.html}.
Ground-based observations revealed a characteristic red color for this object:
according to \cite{boehnhardt2014}, its $V-R$ color index is $0.86\pm0.02$.
\cite{santossanz2012} have studied 15 scattered disk objects (SDOs)
and detached objects, including \ortiz{}, where these objects have a series of
far-infrared thermal measurements taken with the {\it Herschel} Space 
Observatory \footnote{{\it Herschel} is an ESA space observatory with science 
instruments provided by European-led Principal Investigator consortia 
and with important participation from NASA.}.
The albedo of \ortiz{} was found to be $p_{R}\approx18\%$ in $R$
band, hence this object is a member of the ``bright \& red'' subgroup
of the TNO population \citep{lacerda2014}. The corresponding diameter
of \ortiz{} was reported as $d=1280\pm210\,{\rm km}$ 
\citep[see also Table 5 in][]{santossanz2012}. The analysis of near-infrared
spectra also revealed the presence of water ice absorption features
\citep{brown2011}.

The {\it Kepler} space telescope has been designed to continuously 
observe a dedicated field close to the northern pole of the Ecliptic
in order to discover and characterize transiting extrasolar planets 
\citep{borucki2010}. After the failure of the reaction wheels, having only
two available for fine attitude control, the new mission called K2 
has been initiated and commissioned \citep{howell2014}. In this extended
mission, {\it Kepler} observes fields close to the ecliptic plane 
in a quarterly schedule. Due to the orientation of the solar panels 
on {\it Kepler}, these fields have a typical solar elongation between 
$\sim 140 - 50$ degrees during such a $\sim 3$ months long campaign. 

Observing near the ecliptic has two relevant consequences. First, minor 
planets crossing the fields could seriously affect the 
photometric quality by intersecting the apertures of target stars
\citep{szabo2015}. Second, allocating dedicated pixel masks to these
moving Solar System objects can provide a unique way to gather 
uninterrupted photometric time series. This can further be relevant for TNOs 
where the apparent mean motion is slow: as it has been demonstrated by
\cite{pal2015}, even small stamps having a size of $\sim20\times20$ pixels
could include the arc of a TNO around its stationary point (which is
also observed in a K2 campaign, see the typical solar elongation range
above). To date, the K2 mission has been involved in the
precise detection of rotation light variations of the objects 
\jj{}, \gv{} \citep{pal2015} and Nereid, a satellite of Neptune
\citep{kiss2016}. In this work we extend this sample with \ortizlong{}.

Up to now, no rotational brightness variation has been detected
for \ortiz{}: the upper limit for a light curve amplitude found
by \cite{benecchi2013} is $<0.09\,{\rm mag}$. 
Using K2 observations, 
we present the first detection of optical brightness variations 
of this object, detecting a slow, likely double-peaked rotation with a
corresponding low amplitude light curve. This information is further 
used to characterize the physical properties of the surface of \ortiz{}
by employing thermophysical models. 
In Sec.~\ref{sec:observations}, we describe the observations and data
reduction related to K2 and the re-reduction of 
{\it Herschel}/PACS scan map data. 
In Sec.~\ref{sec:analysis}, we briefly detail the methods used to
analyze the optical light curve. The description of the accurate thermal 
modelling is found in Sec.~\ref{sec:thermal}. In Sec.~\ref{sec:conclusions},
we summarize our results. 

\section{Observations and data reduction}
\label{sec:observations}

\subsection{Kepler/K2 observations and data reduction}
\label{sec:kepler}

{\it Kepler} observed the apparent track of \ortiz{} in K2 Campaign~3
under the Guest Observer Office proposal GO3053. The track has been
covered by two custom aperture masks following the trajectory of the object
with a width of $10-11$ pixels on average. Unfortunately, the apparent
stationary point of the object, viewed from {\it Kepler}, was 
located in the gap between the two CCDs of module \#18
(in fact, in the gap between channels 2 and 3). 

Hence, the first pixel mask covered the first $\sim15$ days of Campaign 3 while
the second pixel mask covered only the last $\sim5$ days of the planned
interval. Another unfortunate constellation is the apparent vicinity of  
the bright star 45~Aquarii (HD~211676), which has a brightness of $V=5.9$.
The systematics induced by the halo and the diffraction spikes of 
45~Aquarii significantly decrease the attainable signal-to-noise ratio even in
the case of a moving object. However, Campaign 3 ended prematurely   
after $69.2$ days, about $10$ days short of the planned length of the campaign,
therefore \ortiz{} did not appear in the mask closer to 45~Aqr at all
\citep{thompson2015}. Overall, {\it Kepler} followed the light variations 
of \ortiz{} for 12.0 days continuously. 
The elongation of the object decreased from $140$ to $70$ degrees during 
the campaign but due to the aforementioned facts, only the elongations 
between $140$ and $123$ degrees were available for further analysis.

The data series for the track of 
\ortiz{} as well as the comparison stars has a timing cadence corresponding
to K2 long-cadence mode, i.e. $0.0204\,{\rm d}$ (approximately
$29.4$\,minutes). 
These long-cadence stamps are summed from $270$ individual exposures
onboard (in order to save telemetric bandwidth). Each exposure
has a net (useful) integration time of $6.02\,{\rm sec}$, while $\sim8\%$ 
of the time is spent by readout
\citep[see also][for more details]{gilliland2010}.

\begin{figure}
\plotone{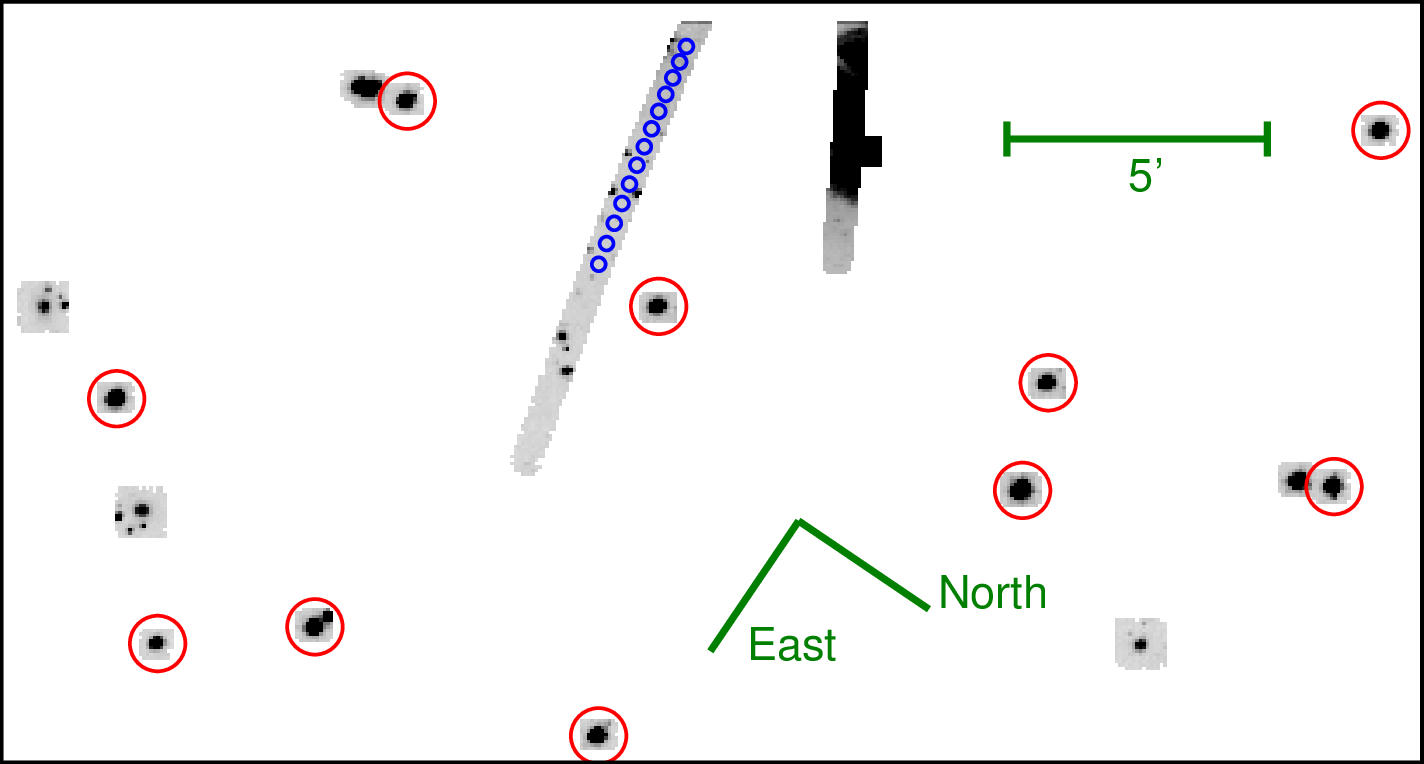}
\caption{The total analyzed field-of-view of 
the {\it Kepler}, showing both the stamps related to \ortizlong{}
as well as the nearby image stamps used for astrometry. The stars used
by the determination for both the differential and absolute
astrometric solutions are indicated by red circles. The field 
has a pixel dimension of $410\times220$, equivalent to
$27^\prime\times15^\prime$. Note that the pixels are shown in the reference
frame of the detector and therefore the image itself is flipped.
Note also that the edge of channel 2 of module \#18 is at the top of the image.}
\label{fig:or10field}
\end{figure}

The public target pixel time series files from the Campaign 3 fields were retrieved 
from the MAST archive\footnote{https://archive.stsci.edu/k2/} for the 
respective observations. In addition to the two masks corresponding to the
parts of the sky covering the apparent arc of \ortiz{}, we retrieved 
a dozen of masks related to nearby additional sources. The analyzed
field-of-view of module \#18 channel 2 has been displayed in
Fig.~\ref{fig:or10field}. Since the masks corresponding to the apparent
trajectory of \ortiz{} do not contain bright background stars, we used 
the information provided by 10 of the unsaturated 
point sources presented on these additional masks to obtain
a relative (differential) and absolute astrometric solutions needed 
by the photometric pipeline. In this sense, this type of astrometric
bootstrapping was simpler than the case of \jjs{} where only the stars
located in the mask corresponding to the object's path were used
\citep[see][for further details]{pal2015}.

The analysis of the frames has been performed in a highly similar manner 
as it was done in the previous K2 observations
\citep{pal2015}. The most relevant improvement in our pipeline
is the inclusion of the aforementioned 
10 additional stamps which provide a more accurate astrometric reference system
w.r.t. the {\it Kepler} CCDs. For all of the processing steps, including the
extraction of K2 data files, we involved the tasks of the
FITSH package\footnote{http://fitsh.szofi.net/} \citep{pal2012}.
As in our previous work \citep{pal2015}, instrumental magnitudes were
derived using differential photometry which is a relatively easy task 
for moving objects when the instrumental point-spread function is stable. 
Individual differential points had a formal uncertainty of $0.07-0.10$\,mags on
average, corresponding to a signal-to-noise ratio of $10-14$. 
This is
in the range of our expectations considering both moving 
objects \citep{pal2015} and faint stationary objects in the
brightness regime of $\sim 21$\,mags  in the original and K2 missions
\citep{molnar2015,olling2015}. 

\begin{deluxetable}{crr}
\tabletypesize{\scriptsize}
\tablecolumns{8}
\tablewidth{0pc}
\tablecaption{Photometric data of \ortiz{}.\label{table:phot}}
\tablehead{\colhead{Time (JD)} & \colhead{Magnitude\ensuremath{^{\rm a}}} & \colhead{Error}}
\startdata
2456982.00186 & 20.942 & 0.087 \\
2456982.02229 & 20.951 & 0.080 \\
2456982.04272 & 20.900 & 0.075
\enddata
\tablecomments{Table~\ref{table:phot} is published in its entirety in the
electronic edition of the {\it Astronomical Journal}.  A portion is
shown here for guidance regarding its form and content.}
\tablenotetext{a}{Magnitudes shown here are transformed to USNO-B1.0 $R$
system, see text for further details.}
\end{deluxetable}

The photometric magnitudes of \ortiz{} have been transformed into USNO-B1.0 
$R$ system \citep{monet2003}. 
In order to find the transformation coefficients, 
we fitted 10 of the additional stars included in the analysis 
(originally selected for astrometric purposes).
We found that the unbiased residual of the 
photometric transformation between USNO-B1.0 and
{\it Kepler} unfiltered magnitudes was $0.09\,{\rm mag}$.
The magnitude of these stars used for this transformation were in the 
range of $R=11$ and $R=14$ (i.e. somewhat brighter regime what was used in
the case of \jj{} earlier). 

We note here that the intrinsic red color of \ortiz{} and
the unfiltered nature of {\it Kepler} observations make this 
type of transformation and hence the yielded magnitudes not be suitable for
physical interpretation. Indeed, the absolute magnitude of \ortiz{}
in $R$ band (see Sec.~\ref{sec:thermal} later on) combined with the
observation geometry at the time of the usable K2 observations
yields an expected $R$ magnitude of $20.88$ while the median of the light curve
is $21.17$ magnitudes. This difference of $\sim 0.3$ magnitudes is significantly
larger than the residual of the photometric transformation and even large
to be accounted for phase effects. The photometric time series data of
\ortiz{} are shown in Table~\ref{table:phot} (the full table is available
in an electronic form). In order to reject the outlier points, we 
performed an iterative sigma-clipping procedure in the binned light curves.
This procedure has significantly decreased the light curve RMS, showing
that these outlier points were caused by non-Gaussian random effects 
(systematics on the detector, cosmic hits, etc). 

The photometric quality can easily be quantified as follows. If one
has a time series of magnitudes and their respective uncertainties
(as derived by the photometric pipeline run on each image separately),
then one can compare the model fit residuals w.r.t these uncertainties.
In our case, we consider the folded and binned light curve as a ``model fit''
If the ratio of these two numbers are close to unity, it means that the 
photometric quality (i.e. the overall efficiency of the photometric pipeline) 
is nearly perfect -- independently of the actual values 
of the uncertainties. In our case, these values are $\sim0.11$ (the mean of
RMS around the binned points) and $\sim0.08$ (the mean value of the photometric
uncertainties as reported by FITSH/\texttt{fiphot}). It means that the 
photometric quality can be considered adequate but indeed there could be 
options to further tune in the algorithms. It can even mean the more 
sophisticated rejection of outliers (due to cosmic hits or prominent residual 
structures on the differential images, etc) could further push this ratio down 
to or at least, closer to unity. 
We note here that this ratio of 
$0.11/0.08\approx1.4$ is even better what was found in the case of \jj{}, 
where it is $\sim1.7$ or what was found for Nereid ($\sim1.8$) but 
worse than what can be derived for \gv{} for which it is $\sim1.1$. 
These comparison can also be done using
the publicly available data for these three objects \citep{pal2015,kiss2016}.
We also note that the similar median stacking procedure which was
involved during the analysis of \gv{} cannot be applied for these 
K2 observations of \ortiz{} since the apparent mean motion
was much higher (i.e. \gv{} was observed during its stationary point
while images for \ortiz{} are available only at the beginning of the campaign,
far off the stationary point). By increasing the sample of photometric
data series of moving objects acquired by K2, we could provide
algorithms which would yield more precise light curves. 
According to the current sample of three such observations, the
respective light curve of \ortiz{} has an ``average quality'' in this sense. 

\begin{deluxetable}{llr}
\tabletypesize{\scriptsize}
\tablecolumns{8}
\tablewidth{0pc}
\tablecaption{Orbital and optical data for \ortiz{}.\label{table:auxdata}}
\tablehead{\colhead{Quantity} & \colhead{Symbol} & \colhead{Value}}
\startdata
Heliocentric distance		& $r$		& $86.331$\,AU		\\
Distance from {\it Herschel}	& $\Delta$	& $86.586$\,AU		\\
Phase angle			& $\alpha$	& $0.\!\!^\circ65$	\\
Absolute visual magnitude	& $H_V$		& $\hvmag\pm\hverr$	\\
Absolute $R$ magnitude		& $H_R$		& $\hrmag\pm\hrerr$
\enddata
\tablecomments{The above data are for the midpoint 
of {\it Herschel} observations, i.e. 2011 May 8. These parameters were
incorporated throughout the thermal analysis. }
\end{deluxetable}

\subsection{Herschel/PACS observations and data reduction}
\label{sec:herschel}

In the framework of the ``TNO's are Cool!'' Open Time Key Programme
\citep{muller2009} of the Herschel Space Observatory 
\citep{pilbratt2010}, the object \ortiz{} has been observed in a similar 
fashion like the another $130+$ trans-Neptunian targets of this project
\citep{kiss2014}. 
The aim was to employ the Photoconductor Array Camera and 
Spectrometer \citep[PACS,][]{poglitsch2010} instrument of {\it Herschel} to
provide thermal flux estimations for these objects in the wavelength
range of $60-210\,{\rm\mu m}$. Since the expected temperature of 
a trans-Neptunian object is in the range of few tens of Kelvins, 
the PACS instrument provides an efficient way to characterize the 
thermal radiation of these bodies. Once the 
thermal fluxes are known, the combination with the optical absolute 
brightness and rotation period 
yields an unambiguous estimation of the size and albedo.

\begin{figure}
\plotthree{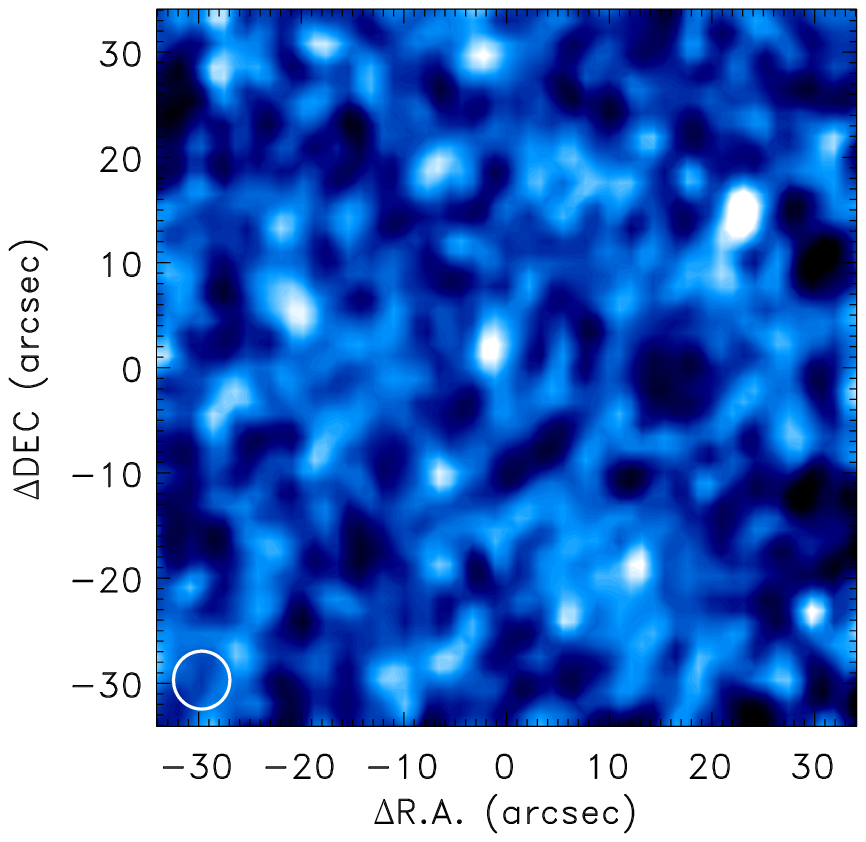}{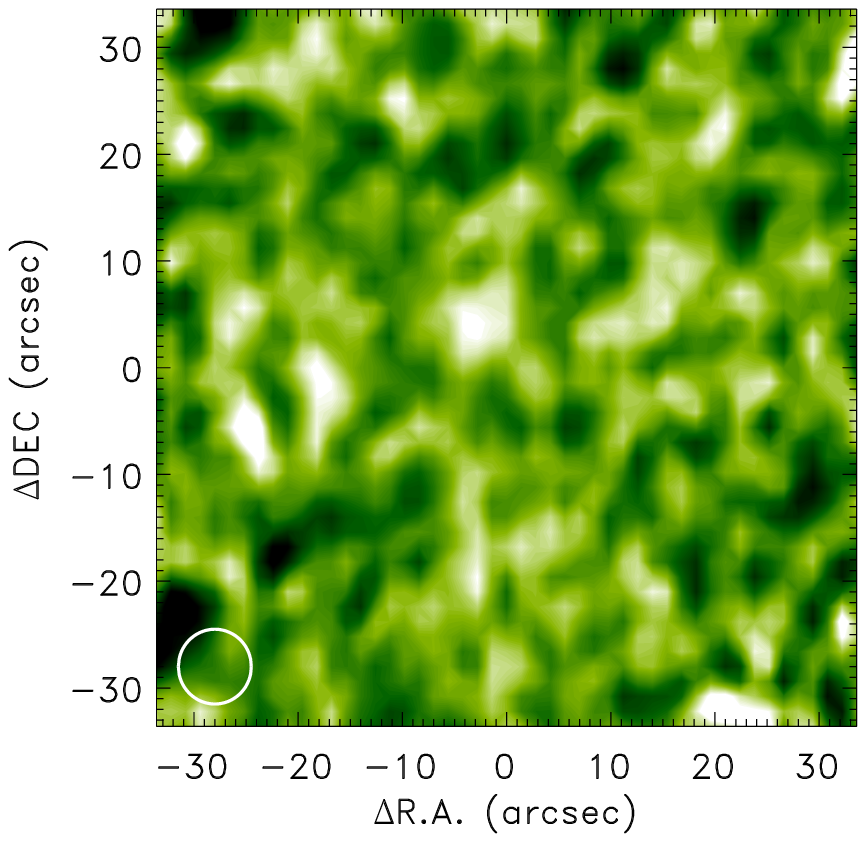}{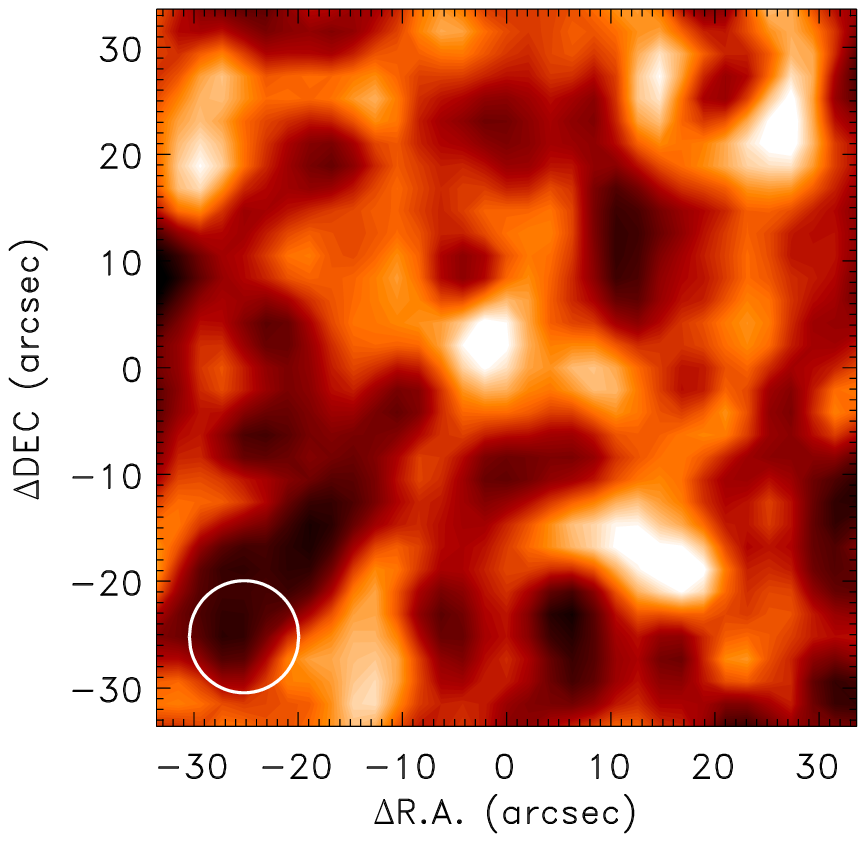}
\caption{Image stamps of \ortizlong{} as seen by the PACS detector 
of {\it Herschel}. The stamps show the vicinity of the object and cover a 
$70^{\prime\prime}\times70^{\prime\prime}$ area on the sky.
From left to right, the stamps show the object in $70\,{\rm\mu m}$ (blue),
$100\,{\rm\mu m}$ (green) and $160\,{\rm\mu m}$ (red) channels. The
small white circles in the lower-left corner show the beam size (which is
the largest in the red channel due to the diffraction-limited resolution
of the instrument. Note that the object itself is slightly offset by
$\approx 2^{\prime\prime}$ from the field center due to the pointing 
drifts and astrometric uncertainties with respect to the nominal 
coordinates.} \label{fig:herschelstamps}
\end{figure}

In brief, a TNO, like \ortiz{} has been observed twice in order to both 
estimate and reduce the effects of the background confusion noise. This
is an essential step since the structure of the background is unknown
due to the lack of any former or recent survey providing imaging data
in this wavelength regime. The summary of {\it Herschel}/PACS observations 
is shown in Table~2 of \cite{santossanz2012}.
Earlier flux estimations have been performed and presented in 
\cite{santossanz2012} for 15 scattered disk and detached objects, 
including \ortiz{}. However, we re-reduced the available 
{\it Herschel}/PACS data using the recent improvements in our HIPE-based
\citep{ott2010} PACS data processing pipeline, presented in \cite{kiss2014}. 
This type of re-reduction involved not only the objects directly related
to the ``TNO's are Cool!'' programme, but exploited additional
observations of recently discovered Solar System targets
\citep[see e.g.][]{pal2015b}. The image stamps created by
this so-called double-differential method \citep{kiss2014,pal2015b} 
are displayed in Fig.~\ref{fig:herschelstamps}. 

Flux estimations have been performed using aperture photometry while 
the respective uncertainties have been derived using the artificial
source implantation method \citep[see also]{kiss2014}. The derived 
uncertainties also include the additional $5\%$ due to the absolute 
flux level calibration error \citep{balog2014}.
The fluxes have been found to be $\bflux\pm\berr\,{\rm mJy}$, 
$\gflux\pm\gerr\,{\rm mJy}$ and $\rflux\pm\rerr\,{\rm mJy}$ in the 
``blue'' ($60-85\,{\rm\mu m}$, centered at $70\,{\rm\mu m}$), 
``green'' ($85-130\,{\rm\mu m}$, centered at $100\,{\rm\mu m}$) and 
``red'' ($130-210\,{\rm\mu m}$, centered at $160\,{\rm\mu m}$) PACS bands. During the derivation
of these fluxes, we also included the color correction factors
of $C_{70}=0.992$, $C_{100}=0.985$ and $C_{160}=0.995$ corresponding
to the temperature of $\sim 37\,{\rm K}$ for this object 
\citep[see also][]{muller2011}.

\section{Optical light curve analysis}
\label{sec:analysis}

\begin{figure*}
\plottwo{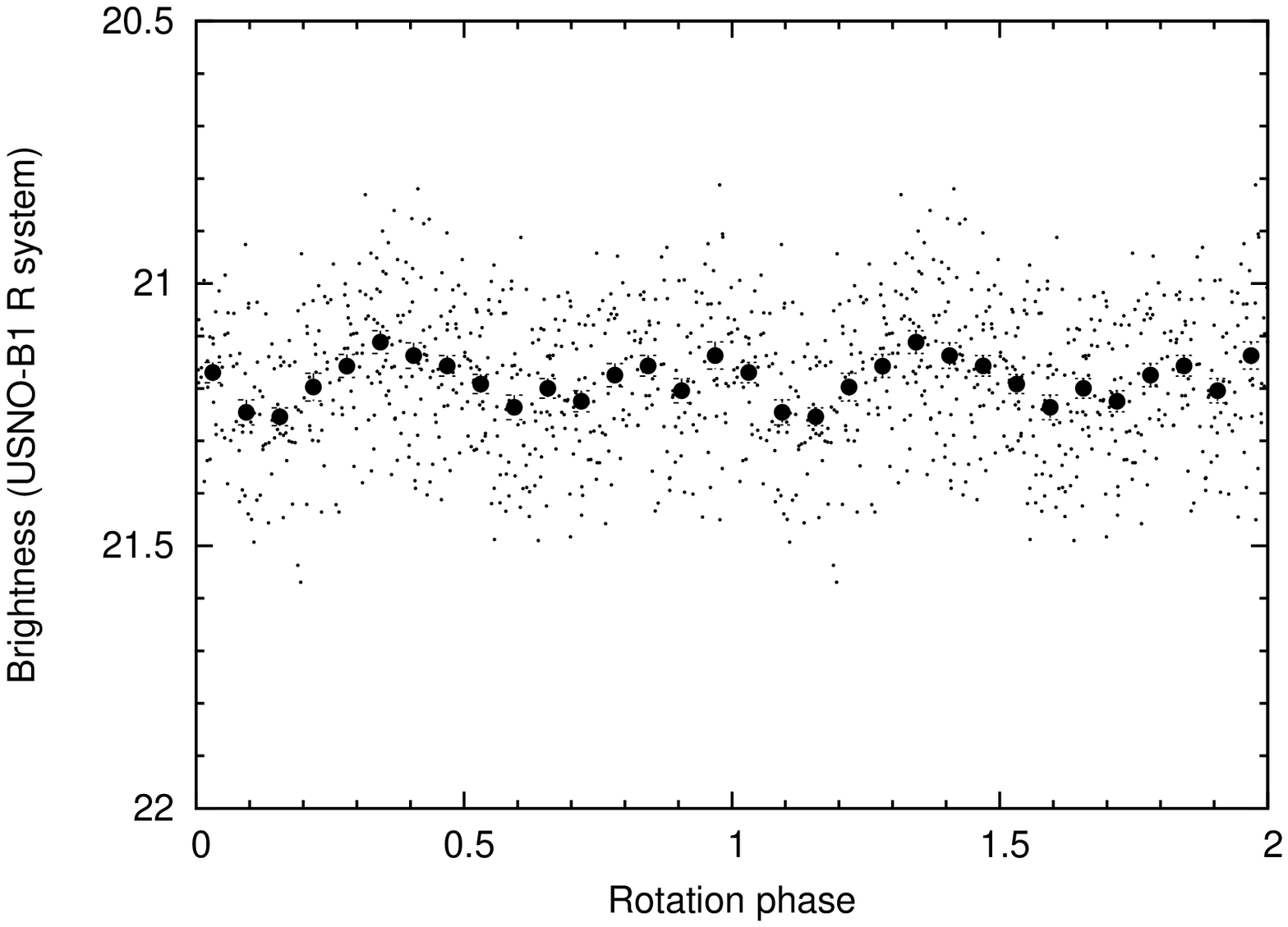}{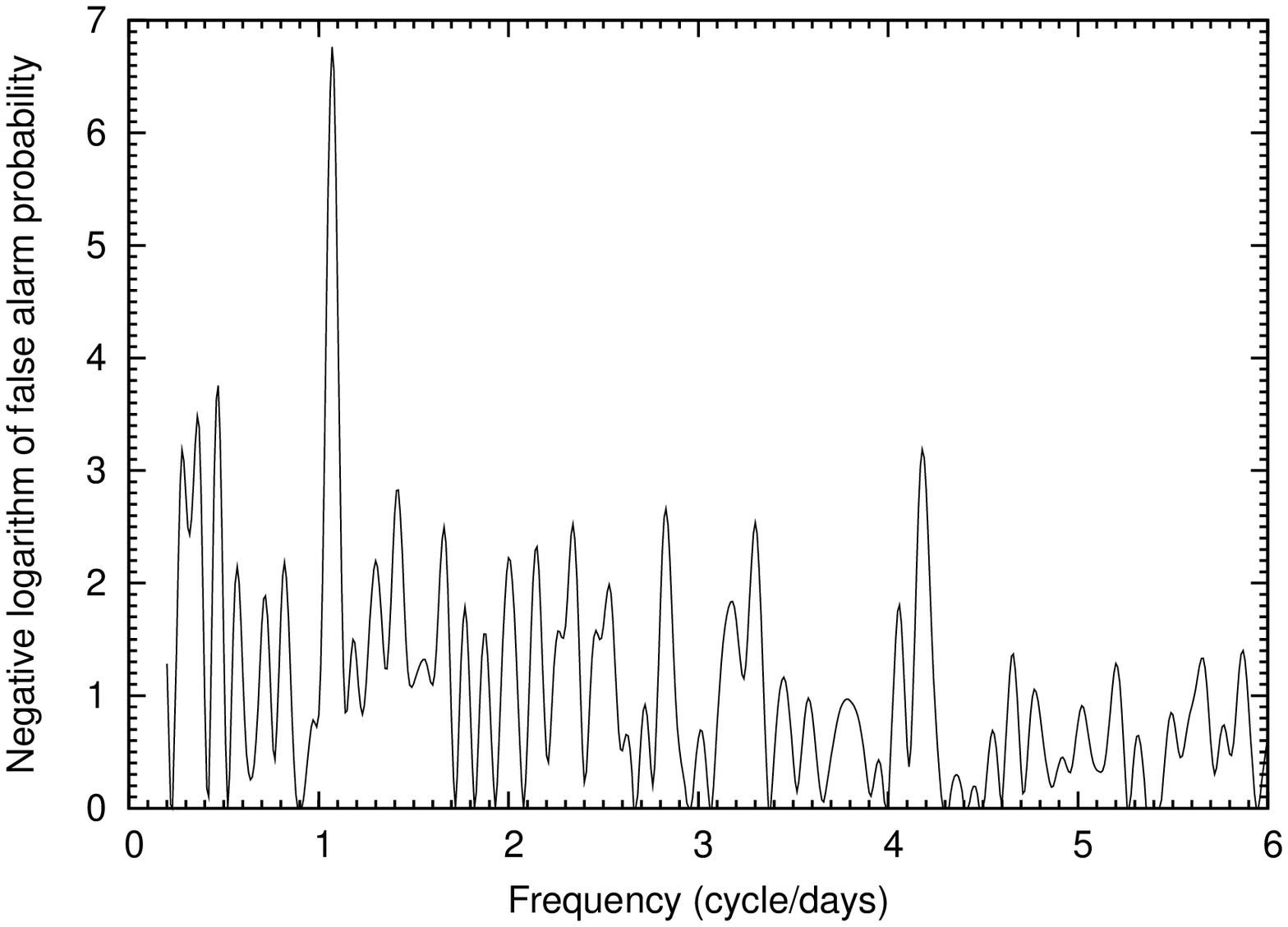}
\caption{Left: phase-folded light curve of 
\ortizlong{} superimposed with binned data points and the
best-fit sinusoidal fit used for period search. The folding period
corresponds to the suspected double-peak rotation period of 
$P_{\rm rot}=44.81\pm0.37\,{\rm h}$. 
Right: Fourier transform of the photometric light variation of \ortizlong{},
as converted to false alarm probabilities, showing the prominent peak at 
$n=1.071\,{\rm cycles/day}$ and the respective false alarm probability of
$1.7\times10^{-7}$. This value corresponds to a detection of $5.2$-$\sigma$.}
\label{fig:lc}
\end{figure*}

In order to find periodicity in the observed K2 photometric time series,
we analyzed the light curve with the Period04 software \citep{lenz2005}.
The Fourier transform of the data revealed a single periodicity
with a signal-to-noise ratio higher than 5.0, at 
$n=1.071 \pm 0.009\,\mathrm{d}^{-1}$. Other peaks, including the frequency 
of the attitude tweak maneuvers, were not detectable in the Fourier spectrum. 
We plot the corresponding false alarm probabilities (in negative log scale)
in the right panel of Fig.~\ref{fig:lc}. We repeated this period search by
fitting a function in a form of 
\begin{equation}
A+B\cos(2\pi n\Delta t)+C\sin(2\pi n\Delta t).\label{eq:freqscan}
\end{equation}
Here $n$ is the scanned rotational frequency and $\Delta t=t-T$,
where $T=2,456,987\,{\rm JD}$ (the approximate center of the time series,
it is subtracted in order to minimize numerical errors). For each frequency
$n$, the unknowns $A$, $B$ and $C$ can be derived using a purely linear
manner. If one converts the fit residuals to false alarm probabilities
(by using the decrement in the corresponding $\chi^2$ values), we got
exactly the same structure what was obtained by Period04. 

Light curves of small Solar System bodies are regularly show double-peaked
features \citep[see e.g.][]{sheppard2007}. 
Therefore, one has to decide whether the the suspected frequency of
$n = 1.071 \,\mathrm{d}^{-1}$ corresponds to a single-peaked light curve
or a light curve having a period which is twice longer. In order to test
the significance of the double-peaked solution, we folded the light
curve with the suspected period of $P_{\rm rot}=44.81\,{\rm h}$ and
performed binning on the folded data series. Using a bin count of $N=16$,
we found that the respective bins differ with a significance of $2.9$-$\sigma$.
This significance is computed as 
\begin{equation}
\sum\limits_{i=0}^{N/2-1} \frac{\left(b_{i+N/2}-b_i\right)^2}{\delta b_{i}^2+\delta b_{i+N/2}^2},\label{eq:assymetrysignificance}
\end{equation}
i.e. by comparing the uncertainty-weighted differences between the 
corresponding bins in the first half of the folded light curve and in the
second half of the folded light curve. If we denote the brightness (magnitude)
in the $i$th bin by $b_i$, then the corresponding binned magnitude
in the next half-phase would be 
$b_{i+N/2}$ (where due to the folding, $b_{i+N}\equiv b_i$, for all integer 
$i$ values). In Eq.~\ref{eq:assymetrysignificance}, $\delta b_{i}$ denotes
the formal uncertainty of the $i$th binned magnitude value. In practice, 
$b_i$ and $\delta b_{i}$ are computed as
\begin{eqnarray}
b_i & = & \frac{\sum\limits_k f_k\Theta\left[i\le\mod(nN(t_k-T),N)<i+1\right]}{B_i}, \\
\delta b_i^2 & = & \frac{\sum\limits_k (f_k-b_i)^2\Theta\left[i\le\mod(nN(t_k-T),N)<i+1\right]}{B_i^2} \nonumber
\end{eqnarray}
where $\Theta(c)$ is unity if the condition $c$ is true, otherwise zero.
Here $\mod(\ell,N)$ is the fractional remainder function (for instance,
$\mod(137.036,42)=11.036$), $k$s are the indices of the light curve points
where the measured magnitude is $f_k$ at the instance $t_k$ and $B_i$ is the number
of points in the $i$th bin, i.e. 
\begin{equation}
B_i=\sum\limits_k\Theta\left[i\le\mod(nN(t_k-T),N)<i+1\right]
\end{equation}
We note here that the above discussed computations can only be done if 
$N$ is even. 

Of course, the value of the significance yielded by Eq.~\ref{eq:assymetrysignificance}
depend on the value of $N$. We found that if we increase the bins
up to $N=20$, $24$ or $32$, we got slightly larger values ($3.0\dots 3.3$).
Hence, this estimate can be considered a conservative one. 
To summarize the above description in brief, we can conclude that the 
probability that the double-peaked solution is preferred against the rotation 
period of $P_{\rm rot}=22.4\,{\rm h}$ is higher than $99\%$. We plot this 
folded and binned light curve on the left panel of Fig.~\ref{fig:lc}. 

In order to further characterize the prominence of the asymmetric two-peaked feature in 
the light curve, we conducted an even more simple procedure. Namely, we
compared the unbiased residuals of the $N=8$ binning against the $N=16$ binning 
points by considering a folding frequency of $n = 1.071 \,\mathrm{d}^{-1}$ 
and $n = 0.535 \,\mathrm{d}^{-1}$, respectively. During the computation
of the unbiased residuals, the degrees of freedom is always the difference between
the light curve points and the number of bins. This comparison yielded 
a $2$-$\sigma$ confidence of the asymmetry in the light curve, and similarly
to the previously described procedure, this value but depends
on the number of bins (yielding confidences in the range of 
$1.5\dots3.0$-$\sigma$). Hence, we can conclude that the true rotation
period is likely corresponding to the double-peaked solution for 
the rotation frequency of $n = 0.535 \,\mathrm{d}^{-1}$ ($P=44.81\,{\rm h}$)
while the single-peaked solution still has a non-negligible chance to
correspond to the true rotation period of $P=22.40\,{\rm h}$. Therefore,
we conduct all further calculations (esp. related to the thermal modelling,
see below) for both possible rotation periods.

By fitting a sinusoidal variation with the aforementioned primary frequency
(by using Eq.~\ref{eq:freqscan}) we found that the respective light curve 
amplitude is $\sqrt{B^2+C^2}=0.0444\pm0.0085$ magnitudes at the frequency peak of 
$n=1.071\,{\rm c/d}$ (see also Fig.~\ref{fig:lc}, right panel).
\cite[by using the tool \texttt{lfit} in the FITSH package, see also][]{pal2012}. 
We note here that this amplitude is compatible with the upper limit 
of $0.09$\,magnitudes found by \cite{benecchi2013}.

As we will see later on (in Sec.~\ref{sec:thermal}), this amplitude
is significantly larger than the uncertainty of the
reported uncertainties of the absolute magnitudes for \ortiz{}
\citep{boehnhardt2014}.
Hence, any formal analysis involving absolute magnitudes must account
for this amplitude as a source for uncertainty since the rotational
phase at the time of the above cited absolute magnitude observations was
practically unknown. Namely, the formal uncertainty of 
$n=1.071\pm0.009\,{\rm c/d}$ is equivalent to  
$1296$ cycles during the timespan between the K2 and
the observations by \cite{boehnhardt2014}, but the total accumulated
error in the rotation phase is $1296\cdot(\Delta n/n)\approx 10.9$. 

\begin{figure*}
\plottwo{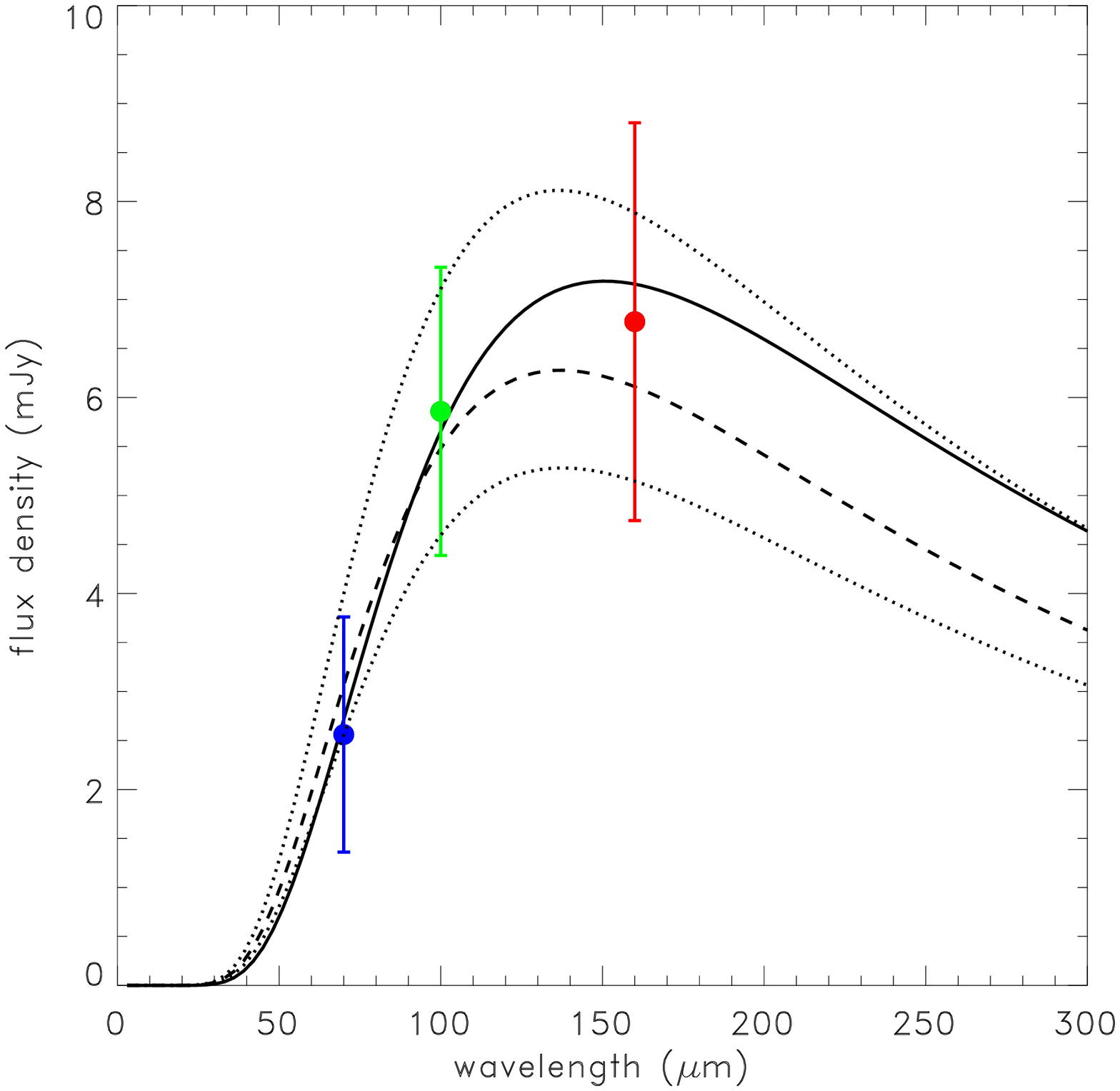}{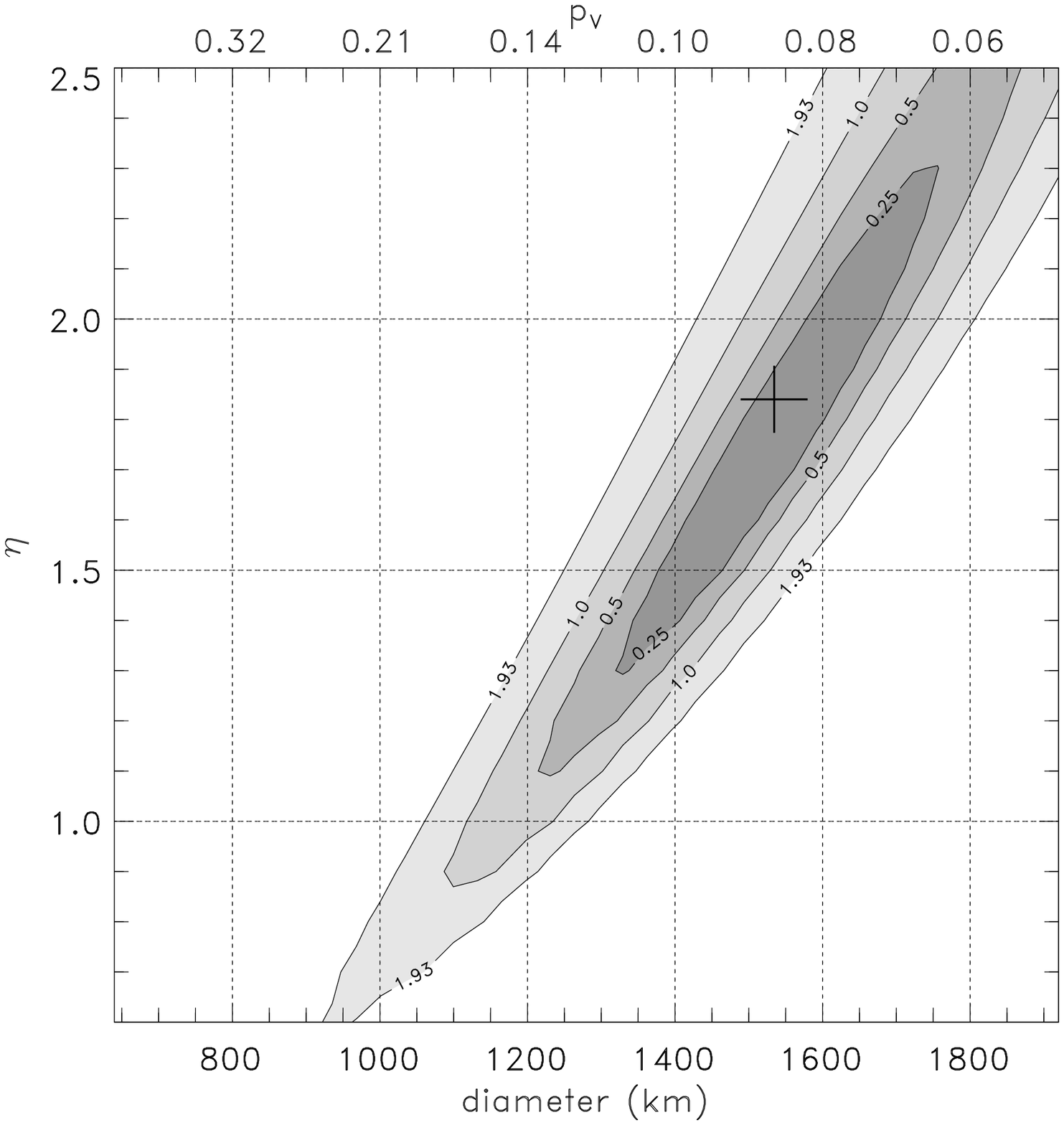}
\caption{Left: measured thermal spectral energy distribution of 
\ortizlong{}, as obtained using {\it Herschel}/PACS measurements.
The thick black curve shows the best-fit Near-Earth Asteroid Thermal
Model (NEATM) curve. The dashed curve represents the best-fit NEATM curve
using a fixed beaming parameter of $\eta=1.2$. The two dotted
curves correspond to the spectral energy distributions defined by the
respective values of diameter and albedo where the value of the
reduced $\chi^2$ is $1.93$. This value of $1.93$ corresponds to the maximum
allowed $\chi^2$ for a two degrees-of-freedom fit. 
Right: contour lines in the reduced $\chi^2$ space as the function of the 
diameter and the beaming parameter in a NEATM fit. The large plus sign 
marks the position of the best fit thermophysical model solution with the 
thermal parameters (thermal inertia and surface roughness) converted 
into beaming parameter.}
\label{fig:or10stm}
\end{figure*}

\section{Thermal modelling}
\label{sec:thermal}

Accurate optical photometry has been carried out by \cite{boehnhardt2014}
in order to derive absolute brightness information of several dozens of 
trans-Neptunian objects which are associated also to the ``TNO's are Cool!''
programme. Their reported absolute magnitudes were $H_{\rm V}=\hvmag\,{\rm mag}$
and $H_{\rm R}=\hrmag\,{\rm mag}$, however, the formal uncertainties given in this work
($0.01$\,mag, in practice, for both $V$ and $R$ colors) are definitely smaller than
the amplitude of the detected light curve variations ($0.0444$\,mag, see above).
Since the rotational phase of this object was unknown at the time of
the corresponding VLT/FORS2 observations, we adopted an additional uncertainty 
in both colors which is equivalent to the amplitude of the light curve 
variations. Namely, in the subsequent thermal modelling we used
$H_{\rm V}=\hvmag\pm\hverr\,{\rm mag}$ and $H_{\rm R}=\hrmag\pm\hrerr\,{\rm mag}$.

\subsection{Near-Earth Asteroid Thermal Model}
\label{sec:stm}

One of the earliest model capable to the computation of 
thermal emission of small Solar System bodies is 
the Standard Thermal Model (STM) by \cite{lebofsky1986}. Basically, this model 
expects a small phase angle for the object and uses an extrapolation
for larger phase angles. However, in the case of \ortiz{}, the phase
angle was quite small at the time of {\it Herschel}/PACS observations
($0.65^\circ$, see also Table~\ref{table:auxdata} for a summary 
of the observation geometry). Hence, this model yields practically the same
results than the sophisticated analysis methods developed for larger phase angles,
such as the Near-Earth Asteroid Thermal Model (NEATM) by \cite{harris1998}. 

Incorporating STM/NEATM in a fitting procedure allows us to obtain the diameter
and geometric albedo of the object by expecting both the thermal fluxes and
the absolute magnitude of the object to be known. 
First, we performed this analysis by involving the aforementioned values of thermal fluxes,
absolute magnitudes and a fixed value of the beaming parameter of $\eta=1.2$
\citep[the mean value of beaming parameters derived by][]{stansberry2008}.
We obtained a diameter of $d=1280^{+130}_{-145},{\rm km}$ and 
$p_V=0.125^{+0.033}_{-0.021}$. By letting the beaming parameter 
$\eta$ to be freely floating during the fit procedure, we got values of
$\eta=1.8\pm0.4$, $d=1550^{+175}_{-190}\,{\rm km}$ and 
$p_V = 0.085^{+0.023}_{-0.016}$. 
We note here that the essential difference between the new estimation
presented in this paper and the one found in \cite{santossanz2012} is 
the treatment of the beaming parameter. While fixing $\eta=1.2$, these
new numbers perfectly agree with that of \cite{santossanz2012}, however,
the derived diameter is certainly larger when we consider the beaming
parameter as an additional free parameter of this type of thermal model.
As we will see later on (in Sec.~\ref{sec:tpm}), more sophisticated
thermophysical models also prefer larger diameters in a nice accordance with
NEATM. 

The spectral energy distribution as well as the corresponding contour lines
in the reduced $\chi^2$ space are displayed in Fig.~\ref{fig:or10stm}. 
The structure of the contour lines imply a strong correlation between
the beaming parameter and the diameter. Due to the lack of 
a more accurate long wavelength thermal flux at $\lambda=160\,{\rm\mu m}$,
the beaming parameter cannot be constrained further (see also the
right panel of Fig.~\ref{fig:or10stm}, where the dashed and solid lines
go very close to each other at $\lambda\lesssim 100\,{\rm\mu m}$).

\begin{figure*}
\begin{center}
\resizebox{55mm}{!}{\includegraphics{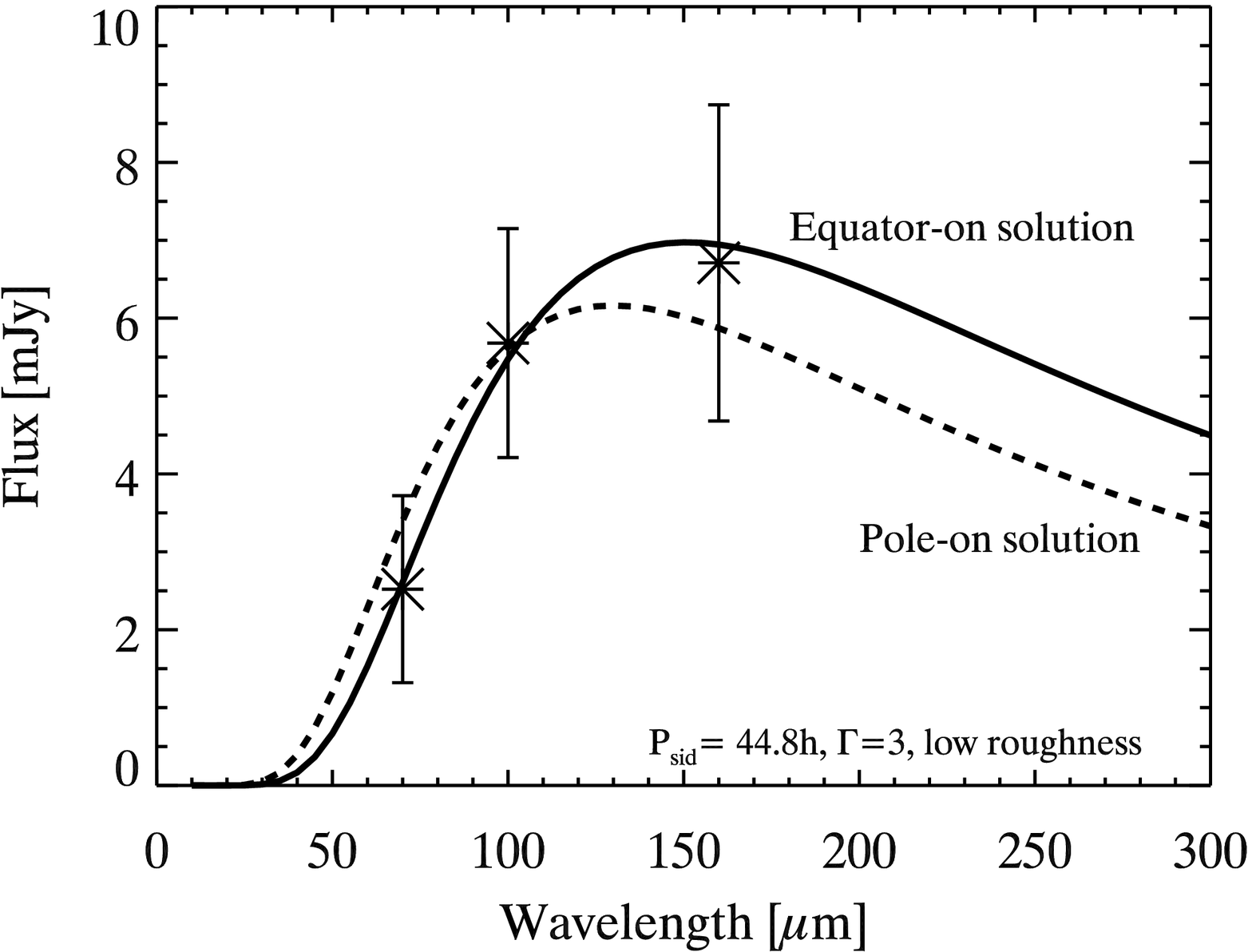}}
\resizebox{55mm}{!}{\includegraphics{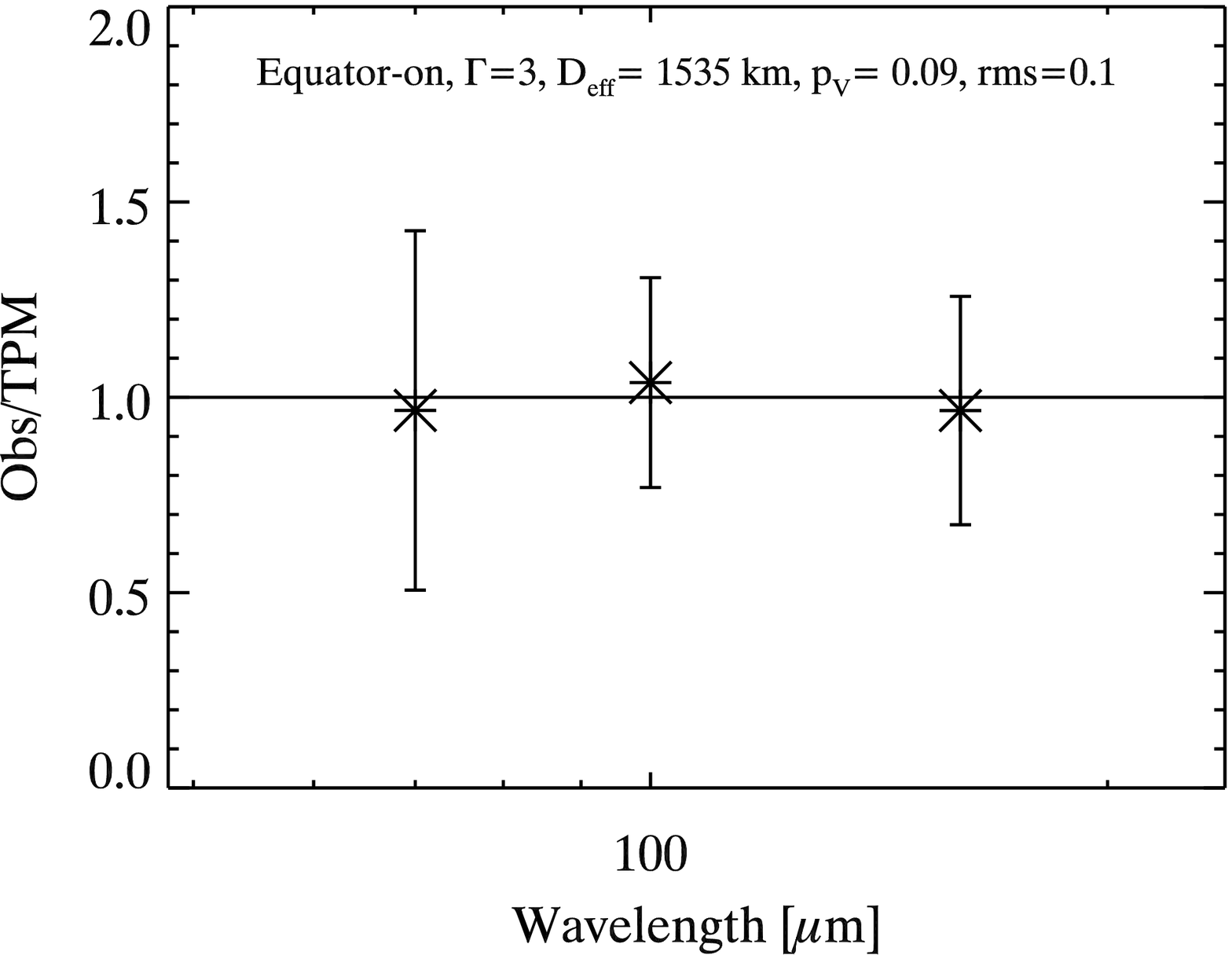}}
\resizebox{55mm}{!}{\includegraphics{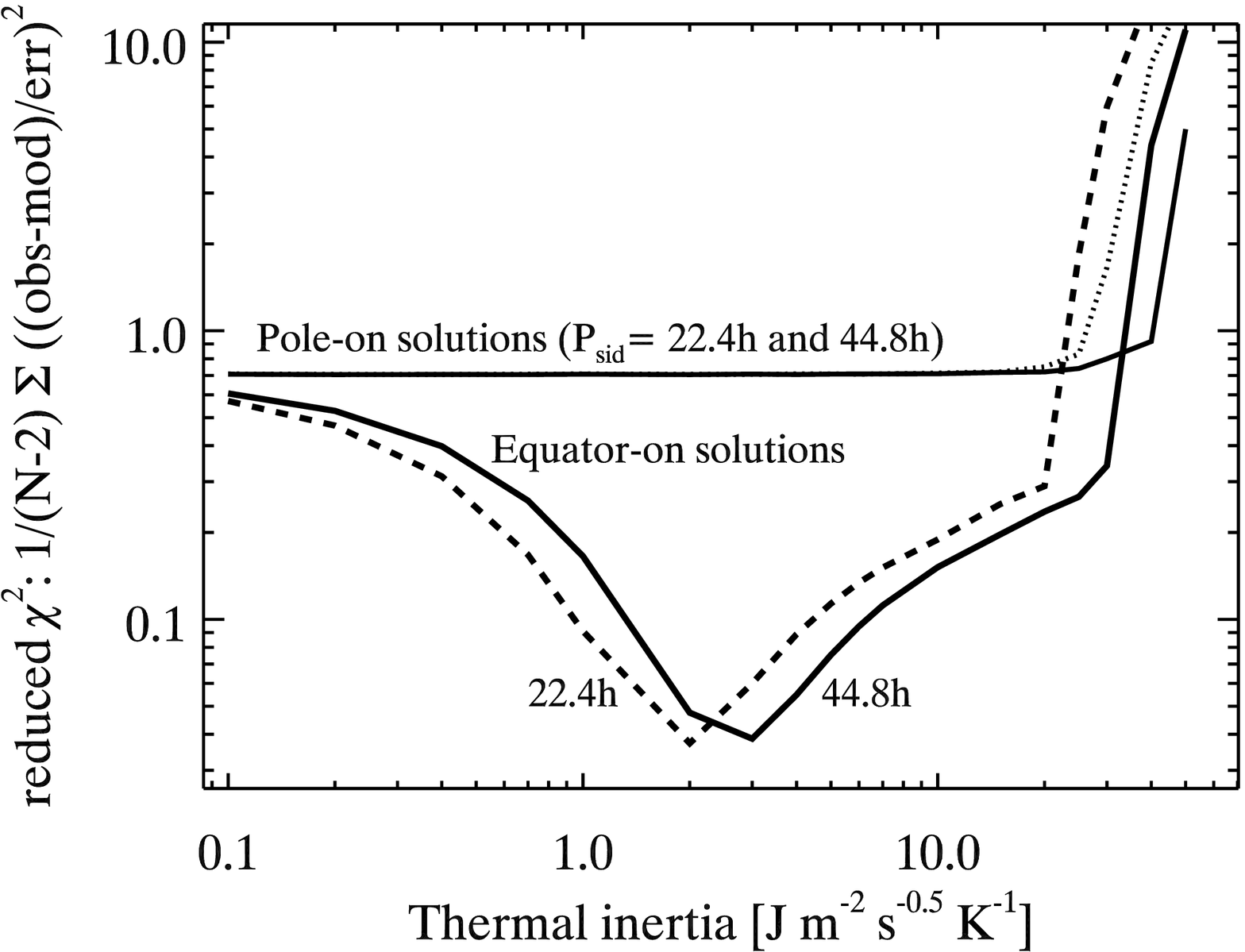}}
\end{center}
\caption{Left: measured thermal spectral energy distribution of 
\ortizlong{}, as obtained using {\it Herschel}/PACS measurements.
The solid curve shows the best-fit TPM model solution corresponding to
an equator-on solution while the dashed curve corresponds to 
the pole-on solution. Middle: the residual flux ratios with respect
to the best-fit equator-on TPM solution. Right:
The value of the reduced $\chi^2$ as the function of the thermal inertia.
The decrement in the value of $\chi^2$ around $\Gamma\approx3\,\tiunit$
is clearly visible for the equator-on configuration
while there is no such feature for the pole-on geometry.
Note that the dashed line shows the $\chi^2$ values for $P_{\rm rot}=22.40\,{\rm h}$
in the case of the right panel. See the text for further details about
the TPM results for the single-peaked rotational period.}
\label{fig:or10tpm}
\end{figure*}

\subsection{Thermophysical model}
\label{sec:tpm}

The thermal emission of a trans-Neptunian object can further be characterized
by involving the asteroid thermophysical model 
\citep[TPM, see][]{lagerros1996,lagerros1997,lagerros1998,muller1998,muller2002}. 
This model incorporates not only the absolute brightness values and the 
thermal fluxes but also the rotation period and the orientation geometry of 
the rotation axis. Throughout our analysis, we tested the possible orientation 
geometries of pole-on, equator-on and zero obliquity with the respective 
$(\lambda,\beta)$ polar ecliptic coordinates of $(331.9,-3.3)$; 
$(331.9, 86.7)$ and $(246.8, 59.2)$.
Our TPM analysis yielded a best-fit solution diameter and albedo close to the 
results of the NEATM fit with free-floating beaming parameter (see above
in Sec.~\ref{sec:stm}). Namely, the best-fit TPM parameters 
for the equator-on geometry and the rotation period of 
$P_{\rm rot}=44.81\,{\rm h}$ are 
$d=1535^{+75}_{-225}\,{\rm km}$ and $p_V=0.089^{+0.031}_{-0.009}$
while the preferred thermal inertia is $\Gamma=3\,\tiunit$. The 
spectral energy distribution along with the measured far infrared
fluxes (corresponding to these model parameters) are displayed in
Fig.~\ref{fig:or10tpm}. 

Strictly speaking, we should note that all of the inertia values of 
$\Gamma \lesssim 20\,\tiunit$ and both equatorial-on and pole-on
geometries provide a consistent fit having $\chi^2\lesssim 1$. 
In other words, PACS data do not allow us to 
constrain the spin-axis orientation, rotation period, thermal inertia or 
roughness. However, the equator-on, as well as the zero obliquity cases
produce more consistent results with reduced $\chi^2$ values well below $1.0$,
see Fig.~\ref{fig:or10tpm}, right panel. 
The aforementioned corresponding value for the thermal inertia 
($\Gamma=3\,\tiunit$) agrees well with the typical thermal inertias for 
very distant TNOs \citep[see][Fig.~13, right panel]{lellouch2013}
which are roughly in the range of $\Gamma=0.7\dots 5,\tiunit$. 
Due to the lower confidence of the double-peaked light curve (see 
Sec.~\ref{sec:analysis}), we repeated the TPM analysis for the same
set of input parameter with the exception of the rotation period which
was fixed to $P_{\rm rot}^\prime=22.40\,{\rm h}$. In this case, we obtained
$d^\prime=1525^{+121}_{-180}\,{\rm km}$ and $p^\prime_V=0.090^{+0.023}_{-0.013}$
while the preferred thermal inertia is $\Gamma^\prime=2\,\tiunit$. These
values differs only marginally from the aforementioned values derived 
for $P_{\rm rot}=44.81\,{\rm h}$. The respective curves are also
shown in the plots of Fig.~\ref{fig:or10tpm}.

In order to be able to compare our NEATM and thermophysical model results, the
thermal parameters of the best fit thermophysical model solution
($d=1535\,{\rm km}$ for the $P=44.81\,{\rm h}$ rotation period and assuming
equator-on geometry) were converted into beaming parameter using the procedure
described in \cite{lellouch2013}, based on the papers by 
\cite{spencer1989} and \cite{spencer1990}. This conversion resulted in a 
beaming parameter of $\eta=1.84$ using $\beta=0^\circ$ subsolar latitude and 
a low surface 
roughness. These best fit diameter and beaming parameter values are in
excellent agreement with the best fit values obtained from the NEATM analysis
(see also the right panel of Fig.~\ref{fig:or10stm}).

\section{Results and conclusions}
\label{sec:conclusions}

Our newly derived diameter of \ortiz{}, $d=1535^{+75}_{-225}\,{\rm km}$
is notably larger than the previously obtained value of \cite{santossanz2012}.
This new value would place \ortiz{} as the third largest 
dwarf planet -- see also Table~3 of \cite{lellouch2013}, after Pluto and Eris. 
Even considering these refined values, this object is a member of the 
``bright \& red'' group of \cite{lacerda2014}.

Due to its large size, \ortiz{} has likely has a shape close to
spherical that may be altered by rotation \citep[see e.g.][]{lineweaver2010}
This should lead to a shape of a MacLaurin spheroid
(semimajor axes $a = b > c$, and a rotation
around the shortest axis) or to a Jacobi ellipsoid in the case of fast rotation
\citep{plummer1919}. 
For a body in hydrostatic equilibrium there is a critical
flattening value, $\varepsilon_{\rm crit}=0.42$, when the shape
bifurcates from a stable MacLaurin ellipsoid solution to a Jacobi ellipsoid
\citep{plummer1919}. This critical value would correspond to a rotation
period of $P=5.7\,{\rm h}$ assuming a density of $1.2\,{\rm g\,cm^{-3}}$ 
\citep[a typical value among trans-Neptunian objects, see. e.g.][]{brown2013}
and higher densities will make this critical rotation period
even shorter. E.g. for a density of $2.5\,{\rm g\,cm^{-3}}$ --
a typical value among dwarf planets \citep{brown2008} -- the rotation period
would just be $3.9\,{\rm h}$ much faster than the rotation period we 
derived for \ortiz{}.

These critical rotation period values are significantly shorter than
either rotation period obtained from K2 observations 
($22.40$ or $44.81\,{\rm h}$) in this present paper. This indicates that the 
rotation curve of \ortiz{} is very likely due to surface albedo variegations.
While the low amplitude variations detected in the
light curve of \ortiz{} can easily be modelled by a single-peaked light curve
and small surface brightness inhomogeneities, the two-peaked solution can also
be modelled with surface brightness variations with significantly larger 
amplitudes. In this case, the surface of \ortiz{} should have areas 
where the albedo varies between $p_V=0.06 \dots 0.12$. These limits
for the albedo values were derived by fitting a surface albedo distribution
characterized by second-order spherical harmonics. 

The slow rotation of \ortiz{} can also be caused by tidal synchronization,
similar to the object \wg{} \citep[see][]{rabinowitz2013} and it was also 
proposed for the objects \gv{} where the slow rotation were first detected
also by K2 \citep[see also][]{pal2015}. By repeating the 
similar calculations like what is in \cite{rabinowitz2013}, we can give
constraints on the separation of the secondary. These calculations
yielded a separation of $\Delta=2.8\times 10^3\,{\rm km}$
or $\Delta=4.5\times 10^3\,{\rm km}$ for the $\sim 22$ and $\sim 44$
hours or rotation periods, respectively -- by expecting two equal-mass bodies 
with an equivalent effective surface and an average 
density of $1.5\,{\rm g/cm^3}$. At the current distance of
\ortiz{}, these separations are equivalent with $0.045^{\prime\prime}$ and
$0.071^{\prime\prime}$, respectively. When considering a mass ratio 
of $8:1$, similar to that of Pluto--Charon system, the separation slightly
increases to $\Delta=3.0\times 10^3\,{\rm km}$ and 
$\Delta=4.8\times 10^3\,{\rm km}$. Of course, a scenario like the 
Eris--Dysnomia system can also be feasible with much significant contrast 
between the surface brightnesses, however, the magnitude of the expected 
separation is going to be in the same range 
\citep[see e.g. Sec.~5.3 of][for the actual numbers]{santossanz2012}.
We note here that according to Kepler's Third Law, $\Delta\propto(m+M)^{1/3}$,
changes in the mass distributions and/or densities affect the separation 
only slightly.

The red color of \ortiz{} is likely to be due to the retain of methane,
as it was proposed by \cite{brown2011}.  In Fig.~1 in \cite{brown2011},
\ortiz{} is nearly placed on the retention lines of ${\rm CH}_4$,
${\rm CO}$ and ${\rm N}_2$. The larger diameter derived in our paper
places this dwarf planet further inside the volatile retaining domain,
making the explanation of the observed spectrum more feasible. 

\vspace*{2mm}

\begin{acknowledgements}
We thank the detailed notes and comments of the anonymous referee concerning to
the fine details of observations and data analysis. 
This project has been supported by the 
Lend\"ulet LP2012-31 and 2009 Young Researchers Program,
the Hungarian OTKA grants K-109276 and K-104607,
the Hungarian National Research, Development and Innovation Office (NKFIH) 
grants K-115709 and PD-116175 
and by City of Szombathely under agreement no. S-11-1027. 
The research leading to these results has received funding from the 
European Community's Seventh Framework Programme (FP7/2007-2013) under 
grant agreements no. 269194 (IRSES/ASK), no. 312844 (SPACEINN), and the ESA 
PECS Contract Nos. 4000110889/14/NL/NDe and 4000109997/13/NL/KML. 
Gy.~M.~Sz., Cs.~K. and L.~M. were supported by the J\'anos Bolyai Research 
Scholarship. Funding for the K2 spacecraft is provided by the NASA Science 
Mission directorate. 
The authors acknowledge the Kepler team for the extra efforts to 
allocate special pixel masks to track moving targets. 
All of the data presented in this paper were obtained from the 
Mikulski Archive for Space Telescopes (MAST). 
STScI is operated by the Association of Universities for Research in 
Astronomy, Inc., under NASA contract NAS5-26555. Support for MAST for 
non-HST data is provided by the NASA Office of Space Science via 
grant NNX13AC07G and by other grants and contracts.
\end{acknowledgements}


{}

\end{document}